\documentstyle[seceq,epsf,twoside]{ptptex}
\notypesetlogo  
\title{Exploring the structure of a possible light scalar nonet}
\author{%
Deirdre {\sc Black}, Amir H. {\sc Fariborz}
and Joseph {\sc Schechter}}
\inst{%
 Department of Physics , Syracuse University\\
Syracuse, NY 13244-1130, USA\\}
\abst{%
We first review the work of the Syracuse group, which uses an effective
chiral Lagrangian approach, on meson-meson scattering. An illustration 
providing evidence for the existence of a strange scalar resonance
of mass around 900 MeV is given. An attempt to fit this $\kappa (900)$
together with a similarly obtained $\sigma (560)$ and the well known
$a_0(980)$ and $f_0(980)$ into a nonet pattern suggests that the
underlying structure is closer to a dual quark-dual antiquark than to a
quark-antiquark. A possible mechanism to explain a next higher-in mass
scalar meson nonet is also discussed. This involves mixing between
$q{\bar q}$ and $qq{\bar q} {\bar q}$ states.}

\begin{document}
\maketitle

\setcounter{tocdepth}{4}

\section{Introduction}
The possible existence of light scalar mesons (with masses less than
about 1 GeV) has been a controversial subject for roughly forty years.
The last few years have seen a revival of interest in this
area. We will discuss two related aspects. The first involves
the determination of the existence of light scalar mesons and their
properties by comparing models for meson-meson scattering with experiment.
Other speakers\cite{otherspeakers} at this workshop have presented various
approaches to this problem. The work of the Syracuse group
\cite{sg1,sg2,sg3}, discussed in the talk of M. Harada for the case of
$\pi \pi$ scattering, is based on an effective non-linear chiral
Lagrangian
containing pseudoscalar, vector and scalar particles. It is well known
that $ \pi\pi$ scattering very near threshold can be accurately treated
with a chiral Lagrangian of only pseudoscalars, which is systematically
expanded to include all terms with a given
number of derivatives (chiral perturbation theory). However this
essentially polynomial expansion can not be used to explain the shape 
of the scalar partial wave amplitude up to the 1 GeV region without using
a prohibitively large number of derivatives.
The inclusion of scalar resonances directly, provides a much more
economical description, already at tree level, over this extended range.
The tree level scattering amplitude obtained from the chiral Lagrangian
is crossing symmetric but has physical divergences at the direct channel  
poles. These are regularized according to the prescription (for
a light, broad resonance like the $\sigma$ or $\kappa$):
\begin{equation}
 \frac{MG}{M^2-s} \rightarrow \frac{MG}{M^2-s-iMG^\prime} ,
\label{reg}
\end{equation}
where $G^\prime$, which is not required to equal $G$, is taken as
a fitting parameter. $G$ and $M$ are parameters from the chiral
Lagrangian. Fitting the resulting amplitude to experiment, of course,
restores unitarity. In this way both unitarity and crossing symmetry are
approximately satisfied. For the $\pi \pi$ case the amplitude up to 1
GeV has four parts: i. "current algebra" contact term, ii. vector meson
exchange terms, iii. $\sigma(560)$ exchange terms, iv. $f_0(980)$ exchange
terms including the appropriate background (Ramsauer Townsend effect). A
similar pattern seems to hold for $\pi K$ scattering as we will briefly
describe in section 2.

    The second aspect we discuss is the underlying quark structure 
of the light scalars which are needed in our treatment of meson-meson
scattering. As examples, three models for the underlying quark structure
have been discussed by many authors: i) the $K{\bar K}$ molecule model 
\cite{kkbar},
ii) the $q{\bar q}$ model with strong meson-meson interactions
(or "unitarized quark model")\cite{uqm}, iii) the intrinsic $ qq{\bar
q}{\bar q}$
model (Jaffe type\cite{jaffe}). These models have the common feature that
four quarks
are involved in some form; all are different from the ``simple"
$q{\bar q}$ model.  Note that in the
effective Lagrangian approach, the quark substructure of the scalars is
not specified. In particular a nonet field can {\it a priori}
represent either $q{\bar q}$ or $qq{\bar q}{\bar q}$ (or even more
complicated) states since both may have the same flavor transformation
property. Information about the quark structure may however be inferred
indirectly.

\section{Pi K scattering}

    The $J=0$ partial wave
amplitudes of $\pi K$ scattering were treated\cite{sg2} in a similar way
to those of $\pi\pi$ scattering.
 In this case the low energy
amplitude is
taken to correspond to the sum of a current algebra contact diagram,
vector $\rho$ and $K^*$ exchange diagrams and scalar $\sigma(550)$,
$f_0(980)$ and $\kappa(900)$ exchange diagrams. The situation in the
interesting $I=1/2$ channel turns out to be very analogous to
the $I=0$ channel of s-wave $\pi\pi$ scattering. Now a $\kappa(900)$
parametrized as in (\ref{reg}) is required to restore unitarity; it
plays the role of the $\sigma(550)$ in the $\pi\pi$ case. Following
our criterion we expect that to extend this treatment to the 1.5 GeV
region, one should include the many possible exchanges of particles
with masses up to about 1.5 GeV. Nevertheless we found that
a satisfactory description of the 1-1.5 GeV s-wave region is
obtained simply by including the well known $K_0^*(1430)$ scalar
resonance, which plays the role of the $f_0(980)$ in the $\pi\pi$
calculation. 

    It may be helpful to give a step by step pictorial approach to
see how the individual components contribute to the real part of the
$I=1/2$ scalar partial wave amplitude, $R_0^{1/2}$ of $\pi K$ scattering.
In Fig. (1) it is seen that the "current algebra" (i.e. contact term
of the non-linear chiral Lagrangian) violates the unitarity bound,
$|R_0^{1/2}| \leq 1/2$, already not too far from threshold. The inclusion
of vector meson exchanges (dashed line) improves the situation a lot
but still leads to unitarity violation. The unitarity bound may be
satisfied if exchanges of the scalar mesons $\sigma (560)$, $f_0(980)$
and a putative $\kappa(900)$ are included, as shown in Fig.(2). The
s-channel pole of the $\kappa(900)$ was modified as in (\ref{reg})
and the corresponding parameters were obtained by fitting to experiment.
Finally the effect of the $K_0^*(1430)$ resonance is included as shown in
Fig.(3). This inclusion was made with the background corresponding to
Fig.(2). Since this approach involves fitting to experiment, unitarity
is obeyed (with the appropriate elasticity assumption) rather than just
the unitarity bound.
\begin{figure}
\centerline{
\epsfxsize=10 cm
\epsfysize=8 cm
\epsffile{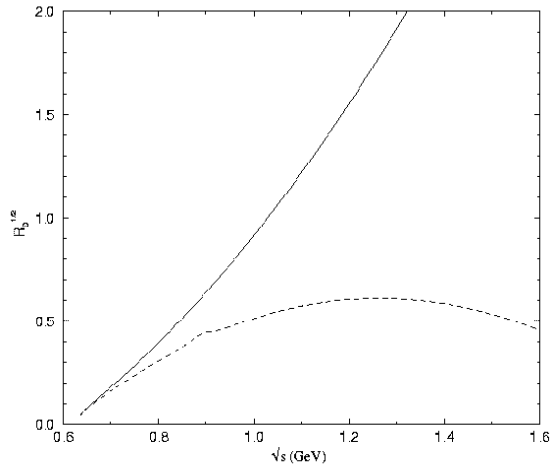}
}
\caption{ Contribution of current algebra (solid line) and current
algebra $+$ vectors (dashed line)to $R_0^{1/2}$.}
\end{figure}
\begin{figure}
\centerline{
\epsfxsize=10 cm
\epsfysize=8 cm
\epsffile{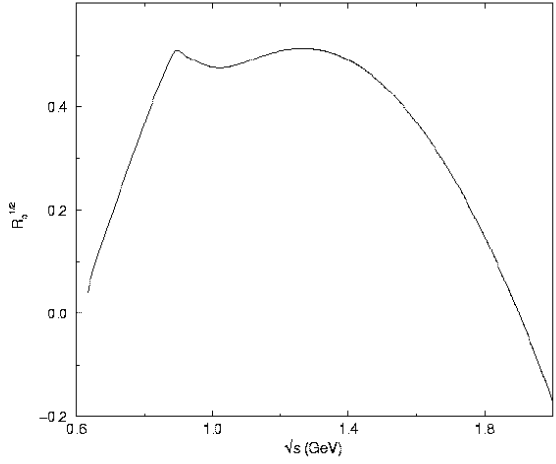}}
\caption{ Contribution of
current algebra $+$ vectors $+ \sigma +f_0(980) + \kappa$ to $R_0^{1/2}$.}
\end{figure}

\begin{figure}
\centerline{
\epsfxsize=10 cm
\epsfysize=8 cm
\epsffile{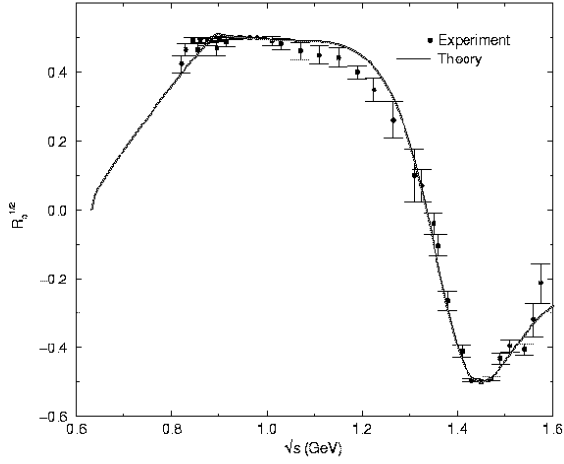}}
\caption{
 Comparison of the theoretical prediction of $R_0^{1/2}$ with the 
experimental data.}
\end{figure}
    It is interesting to observe that our fit to the $R_0^{1/2}$
amplitude has the same general structure as the one used by the
experimentalists in their analysis of the data\cite{aston}. Specifically,
they write the amplitude as the sum of an effective range background piece
and a $K_0^*(1430)$ piece modified by this background. In our model their
background corresponds to the sum of "current algebra", $\rho$,
$\sigma(560)$ exchange, $f_0(980)$ exchange and $\kappa(900)$
exchange pieces. Certainly the effective range description is more
economical. However pieces corresponding to current algebra, vector
meson and at least $f_0(980)$ seem to definitely exist in nature. Our
evidence for the need of a $\kappa(900)$ is in a model in which 
these other contributions are included. If one does not include
these known other contributions, the statistical evidence for
a $\kappa(900)$ would be weaker\cite{cherry}. Our conclusion agrees with
that of Ishida et al\cite{Ishida}.

\section{Scalar nonet ``family" properties}

    The nine states associated with the $\sigma(550)$, $\kappa(900)$,
$f_0(980)$ and $a_0(980)$ are required in order to fit experiment 
in our model. What do their masses and coupling constants suggest
about their quark substructure? (See \cite{putative} for more details.) 
Suppose we first try to assign them
to a conventional $q{\bar q}$ nonet:

\begin{equation}
      \sigma(550)\sim \frac{1}{\sqrt{2}}(u{\bar u} + d{\bar
d}),\,
      \kappa^+(900)\sim u{\bar s},\,
       a_0^+(980)\sim u{\bar d},\,
      f_0(980)\sim s{\bar s} .
\label{conven}
\end{equation}
Then there are two puzzles. i) Why aren't the $a_0(980)$ and the
$\sigma(550)$, which have the same number of non--strange quarks,
degenerate? ii) Why aren't these particles, being p--wave states, in the
same 1+ GeV energy region as the other p--wave states?

    To study this, first note that most meson multiplets can be nicely
understood using the concept of ``ideal mixing". In Okubo's
formulation \cite{okubo}, originally applied to the vector meson
multiplet, the meson fields are grouped into a nonet matrix,
\begin{equation}
N_a^b = \left[
\begin{array} {c c c}
N^1_1&a_0^+&\kappa^+\\
a_0^-&N^2_2&\kappa^0\\
{\bar\kappa}^+&{\bar\kappa}^0&N^3_3\\
\end{array}
\right],
\label{nonet}
\end{equation}
where the particle names have been chosen to fit the scalar mesons.
The two $I=0$ states are the SU(3) singlet, $(N^1_1 + N^2_2 +
N^3_3)/\sqrt{3}$ and the SU(3) octet member, $(N^1_1 + N^2_2 -
2N^3_3)/\sqrt{6}$. Okubo's ansatz for the mass terms was,
\begin{equation}
{\cal L}_{mass} = -a{\rm Tr}(NN) - b{\rm Tr}(NN{\cal M}),
\label{mass}
\end{equation}
where $a>0$ and $b$ are real constants and ${\cal M}=diag(1,1,x)$
(with $x=m_s/m_u$) is the ``spurion" matrix which breaks flavor SU(3)
invariance. With (\ref{nonet}) and (\ref{mass}) the $SU(3)$ singlet
and SU(3) octet isoscalar states mix in such a way (ideal mixing)
that the physical mass eigenstates emerge as $(N^1_1 + N^2_2)/\sqrt{2}$
and $N^3_3$. Furthermore there are two mass relations
\begin{equation}
m^2(a_0)= m^2(\frac{N^1_1 + N^2_2}{\sqrt{2}}),\quad 
m^2(a_0) - m^2(\kappa) = m^2(\kappa) - m^2(N^3_3).
\label{massrelations}
\end{equation}
Note that there are two different solutions depending on the sign of $b$.
If $b>0$ we get Okubo's original case where [with the identifications
$a_0\rightarrow \rho$, $\kappa \rightarrow K^*$, $(N^1_1 + N^2_2)/
\sqrt{2} \rightarrow \omega$ and $N^3_3 \rightarrow \phi$] there
is the conventional ordering
\begin{equation}
m^2(\phi)> m^2(K^*) > m^2(\rho) = m^2(\omega) .
\label{conordering}
\end{equation}
This agrees with counting the number of (heavier) strange quarks
when we identify $N^b_a \sim q_a{\bar q}^b$.

    On the other hand if $b <0$ and we identify $N^3_3 \rightarrow \sigma$
and $(N^1_1 + N^2_2)/\sqrt{2} \rightarrow f_0$, the resulting ordering
would be
 \begin{equation} m^2(f_0) = m^2(a_0) > m^2(\kappa) > m^2(\sigma),
\label{dualordering} \end{equation}
 which is in nice agreement with the
present ``observed" scalar spectrum. But this clearly does not agree with
counting the number of strange quarks while assuming that the scalar
mesons are simple quark anti-quark composites. This unusual ordering will
agree with counting the number of strange quarks if we assume instead that
the scalar mesons are schematically constructed as $N_a^b \sim T_a{\bar
T}^b$ where $ T_a \sim \epsilon_{acd}{\bar q}^c{\bar q}^d$ is a ``dual" 
quark. Specifically
\begin{equation} 
N_a^b \sim T_a{\bar T}^b \sim \left[
\begin{array}{c c c} 
{\bar s}{\bar d}ds&{\bar s}{\bar d}us&{\bar s}{\bar
d}ud \\ 
{\bar s}{\bar u}ds&{\bar s}{\bar u}us&{\bar s}{\bar u}ud \\ 
{\bar
u}{\bar d}ds&{\bar u}{\bar d}us&{\bar u}{\bar d}ud 
\end{array} 
\right]
\label{dualnonet} 
\end{equation}
Note in particular that the light $\sigma
\sim N^3_3$ contains no strange quarks. While this picture seems unusual,
precisely the configuration (\ref{dualnonet}) was found by Jaffe
\cite{jaffe} in the framework of the MIT bag model. The key dynamical
point is that the states in (\ref{dualnonet}) receive (due to the spin and
color spin recoupling coefficients) exceptionally large binding energy 
from the ``hyperfine" piece of the gluon exchange interchange:
\begin{equation}
H_{hf} = - \Delta {\sum}_{i,j}({\bf S}_i\cdot{\bf S}_j)({\bf
F}_i\cdot{\bf
F}_j),
\label{hyperfine}
\end{equation}
wherein the sum goes over all pairs $i,j$ while ${\bf S}_i$ and
${\bf F}_i$ are
respectively the spin and color generators acting on the $i^{th}$
quark or antiquark.

    While the picture above seems close to our expectations it is not
quite right in detail. For example the masses do not exactly obey 
(\ref{massrelations}). Furthermore the simplest model for decay would give
that $f_0 \rightarrow \pi\pi$ vanishes, in contradiction to experiment.
Hence we add the extra mass terms
\begin{equation}
{\cal L}_{mass} = {\rm Eq.}(\ref{mass}) - c{\rm Tr}(N)Tr(N)
-d{\rm Tr}(N){\rm Tr}(N{\cal M}).
\label{fullmass}
\end{equation}
The $c$ and $d$ terms give $f_0-\sigma$ mixing. Now we solve for
$(a,b,c,d)$ in terms of the four masses $m_\sigma=$550 MeV,
$m_\kappa=$900MeV, $m_{a_0}=$983.5 MeV and $m_{f_0}=$980 MeV. The
solution boils down to a quadratic equation for (say) $d$. This gives two
possible values for the mixing angle $\theta_s$ defined by,
\begin{equation}
\left( \begin{array}{c} \sigma\\ f_0 \end{array} \right) = \left(
\begin{array}{c c} {\rm cos} \theta_s & -{\rm sin} \theta_s \\ {\rm sin}
\theta_s & {\rm cos} \theta_s \end{array} \right) \left( \begin{array}{c}
N_3^3 \\ \frac {N_1^1 + N_2^2}{\sqrt 2} \end{array} \right).
\label{sf_mix}
\end{equation}
The solution $\theta_s \approx -90^o$, giving $\sigma \approx (N^1_1
+ N^2_2)/\sqrt{2}$ seems to correspond to restoring the $q{\bar q}$ model
(\ref{conven}) for the scalars once more. The other solution $\theta_s
\approx -20^o$ corresponds to $\sigma$ being mainly $N^3_3$ which was just
noted to be a characteristic signature of the $qq{\bar q}{\bar q}$
model (\ref{dualnonet}). The very existence of these two different 
solutions highlights the fact that by just assuming a flavor
transformation property for the scalars we are not forcing a
particular identification of their underlying quark structure.
Different substructures are naturally associated with different
values of the parameters in the same effective Lagrangian. In
any event, the extra terms in (\ref{fullmass}) have restored the ambiguity
about the scalars' structure. We need more information to decide the
issue. For this purpose we look at the trilinear couplings.

    Using SU(3) invariance we write
\begin{eqnarray}    
{\cal L}_{N\phi \phi} =
&A&{\epsilon}^{abc}{\epsilon}_{def}N_{a}^{d}{\partial_\mu}{\phi}_{b}^{e}
{\partial_\mu}{\phi}_{c}^{f}
+ B {\rm Tr} \left( N \right) {\rm Tr} \left({\partial_\mu}\phi
{\partial_\mu}\phi \right)  \nonumber \\ 
&+&C {\rm Tr} \left( N {\partial_\mu}\phi
\right) {\rm Tr} \left( {\partial_\mu}\phi \right) 
+ D {\rm Tr} \left( N \right) {\rm Tr}
\left({\partial_\mu}\phi \right)  {\rm Tr} \left( {\partial_\mu}\phi
\right),
\label{trilinear}
\end{eqnarray}
where $A,B,C,D$ are four real constants and $\phi$ represents the usual
pseudoscalar nonet matrix. 
The derivatives stem from the 
requirement that (\ref{trilinear}) be the leading part of a chiral
invariant object. If desired, we can rewrite the $A$ term as a linear
combination of the usual Tr$(N\partial_\mu\phi\partial_\mu\phi)$
and the three other terms. The motivation for the form given is
that by itself the $A$ term yields zero for $f_0 \rightarrow \pi\pi$
and $\sigma \rightarrow K{\bar K}$, both of which should vanish in 
a dominant "quark-line rule" picture of a $T{\bar T}$ scalar decaying into
two pseudoscalars. Note that all the
coupling constants which enter into our treatment of $\pi\pi$ and 
$\pi K$ scattering depend on just $A$ and $B$; $C$ and $D$ contribute only
to the decays containing $\eta$ or $\eta^\prime$ in the final state.
For examples of couplings:
\begin{eqnarray}
\gamma_{\kappa K\pi}&=&\gamma_{a_0KK}=-2A, \nonumber \\
\gamma_{\sigma\pi\pi}&=& 2B{\rm sin}
\theta_s-\sqrt{2}(B-A){\rm cos}\theta_s, etc.
\label{examples}
\end{eqnarray}
 
    The mixing angle solution which best fits the couplings needed to
explain the $\pi\pi$ and $\pi K$ scattering turns out to be $\theta_s
\approx -20^o$. Together with a suitable choice of $C$ and $D$, the
interactions involving $\eta$ and $\eta^\prime$ are also consistently
described\cite{sg3}. Thus it seems that our results
point to a picture in which the light scalars are closer to dual quark-
dual antiquark rather than simple quark-antiquark type. Very recently
Achasov \cite{achasov} has argued that new experimental data from
Novosibirsk
on the radiative decay $\phi(1020) \rightarrow \pi^0\eta\gamma$
are better fit with a $qq{\bar q}{\bar q}$ type model of the $a_0(980)$.

\section{Possible mechanism for next lowest-lying scalars}

Of course, the success of the phenomenological quark model suggests
that there exists a nonet of ``conventional" $q{\bar q}$ scalars in
the 1+ GeV range. Let us consider the experimental candidates 
\cite{rpp} for the isovector and isospinor members:
\begin{eqnarray}
&a_0(1450)& : M=1474 \pm 19 {\rm MeV},\hskip.5cm \Gamma=265 \pm 13
{\rm MeV},
\nonumber \\
&K_0^*(1450)& : M=1429 \pm 6 {\rm MeV},\hskip.5cm \Gamma=287 \pm 23
{\rm MeV}.
\nonumber
\end{eqnarray}

On the way to taking these states seriously as members of an ordinary
p-wave nonet we encounter three puzzles. i) The mass of the $a_0^+(1450)$
(presumably a $u{\bar d}$ state is greater than that of the
$K_0^{*+}(1430)$ (presumably a $u{\bar s}$ state). ii) The $a_0(1450)$
and $K_0^*(1430)$ are not less massive than the corresponding p-wave
tensor mesons $a_2(1320)$ and $K_2^*(1430)$, as expected from an
$L\cdot S$ interaction (e.g. $m[\chi_{c2}(1p)]>m[\chi_{c0}(1p)]$).
iii) Assuming the known decay modes $K_0^*(1430) \rightarrow K\pi$
and $a_0(1450) \rightarrow \pi\eta, K{\bar K}, \pi\eta^\prime$
saturate the total widths, we have from SU(3) flavor invariance
that $\Gamma[a_0(1450)] = 1.51 \Gamma[K_0^*(1430)]$. However,
experimentally it is $(0.92 \pm 0.12)\Gamma[K_0(1430)]$ instead.

These puzzles can be simply resolved\cite{bfs} if we assume that 
an ideally mixed heavier $q{\bar q}$ nonet $N^\prime$ in turn mixes with
an ideally mixed $T{\bar T}$ nonet $N$ (as in (\ref{dualnonet})) via
\begin{equation}
{\cal L}^\prime = - \gamma {\rm Tr}(NN^\prime).
\label{nonetmixing}
\end{equation}
This mixing term involves the product of six quark fields in our picture
and is related to the instanton determinant.
 The mechanism is driven by the fact that $m(a_0^\prime)
<m(K_0^\prime)$ while $m(a_0)>m(K_0)$. Here the subscript zero refers
to the unmixed $N$ and $N^\prime$ members. The splittings are summarized
in Fig. 4.

\begin{figure}
\centerline{
\epsfxsize=7cm
\epsfysize=7cm
\epsffile{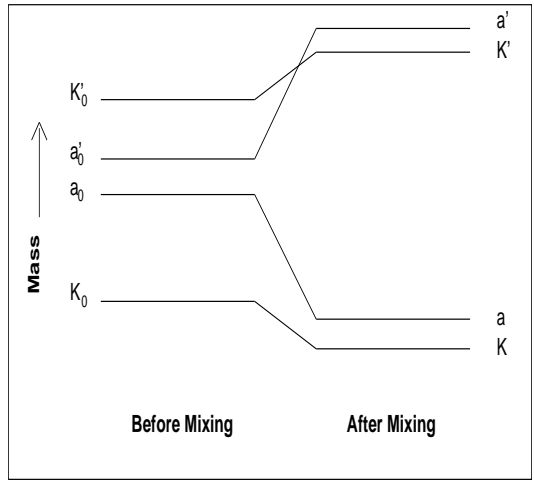}}
\vskip 0.3cm
\caption{Mixing of two nonets-a',K',a and K stand respectively for the
"physical" states $a_0(1450), K_0^*(1430), a_0(980)$ and $\kappa(900)$.
$K_0$ and $a_0$ are the unmixed isospinor and isovector $qq{\bar q}
{\bar q}$ states, while $K_0^\prime$ and $a_0^\prime$ are the
corresponding unmixed $q{\bar q}$ states.}
\end{figure}

The explanations are: i)Think of a perturbation theory approach. There is
a smaller ``energy denominator" for $a_0-a_0^\prime$ mixing than for 
$K_0-K_0^\prime$ mixing. Thus there is more $a_0-a_0^\prime$ repulsion as
shown in Fig. 4. ii) Since the mixing of two levels ``repels" them,
both $a_0(1450)$ and $K_0^*(1430)$ are heavier than would be expected
otherwise. Similarly the light scalars $a_0(980)$ and $\kappa(900)$
are lighter than they would be without the mixing (\ref{nonetmixing}).
iii) The difference between the $a_0(1450)$ and $K_0^*(1430)$
decay coupling constants can be understood from the necessarily greater
mixture of the $qq{\bar q}{\bar q}$ component in the $a_0(1450)$
than in the $K_0^*(1430)$.

This treatment suggests that the light scalar mesons have an interesting
and non-trivial story to tell. Clearly further work will be needed to
complete the picture.

   We would like to thank Masayasu Harada and Francesco Sannino
for fruitful collaboration. One of us (J.S.)
would like to thank the organizers for arranging a stimulating and
enjoyable conference and the Japan Society for the Promotion of Science
for supporting his visit to Japan. 
 The work has been supported in part by the
US DOE under contract DE-FG-02-85ER40231.

\end{document}